\documentclass[%
 reprint,
 amsmath,amssymb,
 aps,
]{revtex4-1}

\usepackage{graphicx}% Include figure files
\usepackage{graphics}
\usepackage{dcolumn}% Align table columns on decimal point
\usepackage{bm}% bold math
\usepackage{float}
\usepackage{caption}
\usepackage{amsmath}
\usepackage{mathrsfs}
\usepackage{amssymb}
\begin{document}

\preprint{APS/123-QED}

\title{Iso-electric point of fluid}
\author{Li Wan}
\affiliation{Department of Physics, Wenzhou University, Wenzhou 325035, P. R. China}
\date{\today}% It is always \today, today,

\begin{abstract}
Iso-electric point(IEP) is the PH, at which the $\zeta$ potential is measured to be zero. The occurrence of IEP has been understood due to the neutralization of surface charge density (SCD) at the solid-liquid interface. In this work, we use the potential trap model to study the sources of the surface charge density at verious PC and PH, by taking the water-silica system as an example. It is revealed that in the case of $PH<8$, the SCD is mainly originated from the dissociation of water molecules. And the bulk ions trapped at the interface can dominate the SCD when $PH>9$. Due to the mass action law, the dissociation of water molecules is suppressed at the PH close to IEP, leading to a zero surface charge density. In this way, zero $\zeta$ potential is obtained at the IEP. It has also been obtained that the increase of the salt concentration in the water can decrease the $\zeta$ potential, but increase the surface charge density. 
\begin{description}
\item[PACS numbers]
47.57.jd, 47.56.+r, 68.08.-p, 66.10.Ed, 73.25.+i
\end{description}
\end{abstract}

%\pacs{66.70.-f,85.80.-g,65.40.gp,72.25.-b,63.21.-e,44.10.+i}% PACS, the Physics and Astronomy
\maketitle

%\tableofcontents

\section{\label{sec1}Introduction}
Electrokinetics(EK) in fluid is a classical subject defined as the study on the interaction of ions with the solid interface, and the dynamics of ions in fluid under external fields~\cite{hunter,journiaux,reppert1,reppert2}. The roots of the EK study are very broad and can be stemmed from various disciplines for decades, such as chemistry, biology, physics, and material sciences. With the development of nano technologies, the structure scale used to confine liquid can be narrowed from micro to nano. The increase of the surface-to-volume ratio in low dimensional structures leads to new EK behaviors~\cite{schoch}. Those new EK phenomena can be applied to molecular manipulation, drug mixing, and electro-osmosis, etc, which stimulates the EK study experiencing a strong revival~\cite{yeh,trau,faure,ramos,green,gonzalez,ajdart,green1,brown,studer,green2,nadal,nadal1,marquet,ristenpart,mpholo,ramos1}.\\

It is well known that the ions at the liquid-solid interface attract counter ions to form electric double layers(EDL) in liquid~\cite{hunter, behrens, davis, sonnefeld,israelachvili,bhatt,dittrich}. As an example of water at $PH=7$, silanol groups at the silica-water interface defined as the stern layer attract $H^+$ ions staying in the neighborhood of the interface by the electrostatic interaction. The EDL is known to play a key role in the EK behaviors. In order to study the charge distribution in the EDL, a conventional method is to solve the Poisson-Boltzmann(PB) equation ~\cite{hunter,ghosal,macdonald,geddes,bazant}. To solve the PB equation, two main types of boundary conditions are adopted. One type boundary condition is called the Gouy-Chapmann(GC) model, in which a surface charge density is assumed at the interface. The other model is to define a $\zeta$ potential as the boundary condition at the slip plane, where the counter ion starts its density extending into the fluid domain. The $\zeta$ potential is an important physical parameter, which reflects the surface charge density and the strength of the EK behaviors. The $\zeta$ potential can be measured indirectly by three main techniques. One is to determine the $\zeta$ potential by measuring electroosmotic mobility combined with the expression of Smoluchowski velocity~\cite{xu, xu1}. The second is by measuring the streaming current or streaming potential~\cite{reppert1}. The last is by measuring electrophoresis behaviors of colloidal particles in liquid~\cite{hidalgo}.\\

In the solving of PB equation, $\zeta$ potential as the boundary condition is considered to be a constant. Similarly in the GC model, the surface charge density of the silanol groups is also considered to be pre-existing constant. The results based on either of the two models are contradictory to the experimental observations~\cite{brian}. In aqueous electrolyte, the $\zeta$ potential and the surface charge density as well are functional of the densities of acid, alkali and salt added into the liquid. For convenience, minus of logarithmic density of $H^+$ ions (salt) is denoted by PH (PC) throughout this paper. At a certain PH, the $\zeta$ potential can be measured to be zero where the PH takes its name of Iso-electric point(IEP). Even though there still has no direct measurements on the surface charge density, the zero $\zeta$ potential at IEP is understood to be attributed to the neutralization of the surface charge density by the acid. The IEP has been observed experimentally for a long time. But it is still lack of a theoretical tool to reveal the occurrence of charge neutralization at IEP. One effective method with a name of charge regulation model(CRM) has been proposed to study the EDL by fitting the experimental data~\cite{behrens1}. However, the CRM has not been used to account for the charge neutralization at IEP.\\

Recently, a surface potential trap model has been proposed to provide a new perspective on the formation of EDL in liquid~\cite{wan}. This model is established in the framework of charge conservation. The surface charge density or the $\zeta$ potential is not fixed as the boundary conditions in the model. They can vary as the PH or PC changes in the liquid. The model has also taken into account the dissociation of water molecules, which has been supported by the reactive molecular dynamics simulation~\cite{fogarty}. In this work, we will use the potential trap model to reveal the mechanism of the charge neutralization at IEP. 

\section{Theory}
A cylindrical channel is considered in this work with a radius of $R$ and an infinite length. In this study, the channel surface is silica and the liquid filled in the channel is water. The acid and alkali are used to adjust the PH and salt is added to change the PC. In such system, $H^+$ and $OH^-$ ions can be not only from the acid and alkali, but also from the dissociation of water molecules. The reactive molecular dynamics simulation shows that on a freshly cut silica surface there exist dangling bonds, Si and Si-O. One neutral water molecule can dissociate into an $OH^{-1}$ and an $H^+$ when contacting the dangling bonds~\cite{fogarty}. The $OH^{-1}$ and $H^+$ ions combine with Si and Si-O to form two silanol groups, which are unstable in an aqueous environment. Silanol groups at the Silica surface contacting water are dissociated, leaving $SiO^{-1}$ ions at the solid surface and $H^+$ in liquid. We skip over the detail of the local chemical reaction at the interface, and propose a surface potential trap to simulate the dangling bonds at the interface physically. The trap height is influenced by the chemical environment of the liquid, such as PC and PH. However, this work will not be focused on the relation between the trap height and the chemical environment, but rather on the mechanism of charge neutralization at IEP. Thus, we will fix the trap height as a constant at each PC. It should be noted that the net charges of $H^+$ and $OH^-$ can be nonzero at $PH \neq 7$, which can be neutralized by the buffer ions. As an example, if $HCl$ is used to decrease the PH of water, then $Cl^-$ is the buffer ion. And the $Na^+$ is the buffer ion if $NaOH$ is used as the alkali. For convenience, the ion densities of $H^+$ and $OH^{-}$ are denoted as $\rho_{w^+}$ and $\rho_{w^-}$ respectively. In this work, we don't distinguish the difference between the buffer ions and the salt ions for simplicity, even though they may have different atomic structures and ionic valencies. The ion densities of the negative and positive ions in the category including the salt and buffer ions are denoted as $\rho_{s^-}$ and $\rho_{s^+}$ respectively. Compared to the  $H^+$ and $OH^{-1}$, the salt and buffer ions are charge conserved, and can not be  produced or eliminated as  $H^+$ and $OH^{-1}$ can do by dissociation or combination in water.\\

The Poisson equation in such system reads:
\begin{equation}
\label{poisson}
\epsilon\nabla^2 \psi = e[\rho_{w^-}+z_1\cdot \rho_{s^-}-\rho_{w^+}-z_2\cdot \rho_{s^+}].
\end{equation} 
with $\epsilon$ the dielectric constant in liquid, $e$ the elementary charge of electron and $\psi$ the potential in the system. $z_1$ and $z_2$ are the ionic valencies for the negative ion and positive ion of salt respectively. Regarding the dissociation of water molecules, the ion densities of $H^+$ and $OH^{-1}$ follow the Boltzmann distribution
\begin{equation}
\begin{split}
&\rho_{w^+}=n^{\infty}\alpha_+\exp^{- e(\psi+f_w-\mu)/(KT)},\\
&\rho_{w^-}=n^{\infty}\alpha_-\exp^{+ e(\psi+f_w-\mu)/(KT)},
\end{split}
\end{equation}
with the chemical potential $\mu$ introduced\cite{wan}. $K$ is the Boltzmann constant and $T$ refers to the room temperature in this work. Here, the potential trap $f_w$ has been implemented in the distribution, and $n^{\infty}$ is equal to $10^{-7}mol/l$ denoting the average ion density in the bulk neutral limit of water at $PH=7$. The parameters of $\alpha_+$ and $\alpha_-$ are used in the equation to exhibit the bulk limit ion density of $H^+$ and $OH^-$ renormalized by the $n^{\infty}$ in the case of $PH \neq 7$. The relevant mass action law for the deprotonation reaction reveals 
\begin{equation}
\begin{split}
&\alpha_+\cdot \alpha_-=1,\\
&\alpha_+=10^{7-PH}. 
\end{split}
\end{equation}
The potential trap takes the following function
\begin{equation}
f_w(r)=\left\{
\begin{array}{lll}
\frac{\gamma}{2}\left(1+\cos \frac{\pi(r-R)}{\Delta}\right), ~&for~&R-\Delta\leq r\leq R\\
0,~&for~&0\leq r\leq R-\Delta.
\end{array}
\right.
\end{equation} 
Here, $\Delta$ is the potential trap width, approximating the length of dangling bond $8\AA$. $\gamma$ is the potential trap height. In this work, we neglect the precise relation between the $\gamma$ and PC or PH, and we use $\gamma=608mV$ at $PC=1, 2, 3$. Since the salt and buffer ions are charge conserved, the $\rho_{s^+}$ and $\rho_{s^-}$ should be governed by
\begin{equation}
\begin{split}
&\rho_{s^+}=\frac{R^2(z_1C_0+\beta_+)\exp^{- ez_2\cdot(\psi+f_s)/(KT)}}{2\int_0^R \exp^{- ez_2\cdot(\psi+f_s)/(KT)}rdr},\\
&\rho_{s^-}=\frac{R^2(z_2C_0+\beta_-)\exp^{+ ez_1\cdot(\psi+f_s)/(KT)}}{2\int_0^R \exp^{+ ez_1\cdot(\psi+f_s)/(KT)}rdr},
\end{split}
\end{equation}
in the cylindrical coordinator~\cite{lc}. Here, $C_0$ is the salt concentration, having the value of $C_0=10^{-PC}mol/l$. The $f_s$ is the potential trap at the liquid-solid interface seen by the salt ions. $\beta_+$ and $\beta_-$ are introduced for the positive and negative buffer ion densities respectively. The buffer ion densities should neutralize the net charges of $H^+$ and $OH^-$, leading to explicit expressions as
\begin{equation}
\begin{split}
&\beta_+=n^{\infty}(\alpha_--\alpha_+)/z_2,\\
&\beta_-=0,
\end{split}
\end{equation} 
for $PH\geq 7$, and
\begin{equation}
\begin{split}
&\beta_+=0,\\
&\beta_-=n^{\infty}(\alpha_+-\alpha_-)/z_1,
\end{split}
\end{equation} 
for $PH\leq 7$. In this study, we use monovalent salt with $z_1=z_2=1$, and set $f_s=0$ for simplicity.\\

Owing to the charge neutralization, the integration of the right hand side of the eq.(\ref{poisson}) over the whole computation domain should be zero, leading to the chemical potential expressed as
\begin{equation}
\mu =\frac{KT}{e}\ln \left[\frac{\alpha_+-\alpha_-+\sqrt{\alpha_+^2+\alpha_-^2-2+4I_+I_-}}{2\alpha_+I_+} \right]
\end{equation}
for both $PH \geq 7$ and $PH \leq 7$ cases. Here, we introduce two integral parameters for convenience
\begin{equation}
\begin{split}
I_+&=\frac{2}{R^2}\int_0^R \exp^{- e(\psi+f_w)/(KT)}r dr,\\
I_-&=\frac{2}{R^2}\int_0^R \exp^{+ e(\psi+f_w)/(KT)}r dr.
\end{split}
\end{equation}
Principally, the charge densities and potential distribution can be solved from the above equations. In the solution, the boundary condition of $\psi=0$ is applied at the interface. After obtaining the $\psi$, the $\zeta$ potential then can be defined generally by
\begin{equation}
\zeta=-\frac{2}{R^2}\int_0^R \psi r dr.
\end{equation}
\\
  
It has been noted that the potential trap can cause the dissociation of water into ions~\cite{wan}. The mount of $H^+$ ($OH^-$) ions contributed from the water dissociation can be obtained from the subtraction of the whole $H^+$ ($OH^-$) ion number in the system by the integral of $n^{\infty}\alpha_+$($n^{\infty}\alpha_-$) over the whole domain. In order to show the dissociation efficiency of molecules, surface charge density related to the dissociation has been defined and noted as $\sigma_1$ and $\sigma_2$ for the $H^+$ and $OH^-$ ions respectively. Due to the charge conservation, $\sigma_1$ and $\sigma_2$ can be solved from the following equations
\begin{equation}
\begin{split}
n^{\infty}\alpha_+I_+\exp^{e\mu/(KT)}=n^{\infty}\alpha_++\frac{2\sigma_1}{R},\\
n^{\infty}\alpha_-I_-\exp^{-e\mu/(KT)}=n^{\infty}\alpha_-+\frac{2\sigma_2}{R}.
\end{split}
\end{equation}
to get
\begin{equation}
\sigma_1=\sigma_2=\frac{n^{\infty} R\alpha_+}{2}[I_+\exp^{e\mu/(KT)}-1].
\end{equation}
The equality of $\sigma_1=\sigma_2$ is consistent to the dissociation process of water molecules in which the dissociated charges of $H^+$ and $OH^-$ ions should be equal to each other. In the following, we will use one parameter $\sigma$ to replace the two denotations of $\sigma_1$ and $\sigma_2$ since $\sigma_1=\sigma_2$.\\

The total surface charge density of $OH^-$ trapped at the interface is denoted by $s$, which represents the integral of $\rho_{w^-}$ in the potential trap
\begin{equation}
s=\frac{1}{R}\int_{R-\Delta}^R \rho_{w^-}rdr.
\end{equation}
It should be noted that the salt ions also can enter into the potential trap at large PC. However, such mount of salt ions in the trap make little influence on the $\zeta$ potential and the obtained results. Thus, we consider the stern layer comprising of only $OH^-$ and exclude the salt ions out of the trap. Another important surface charge density denoted as $b$ is related to the $OH^-$ ions density in bulk limit. Suppose all the $OH^-$ ions in the bulk limit liquid are trapped at the interface, then $b$ is defined as
\begin{equation}
b=n^{\infty}\alpha_-R/2.
\end{equation}  

\section{results}
The cylindrical channel used for the calculation is with $R=20\mu m$ to avoid the overlap of Debye layer.  In the calculation, the trap height is $\gamma=608mV$ for $PC=1,2,3$. Results presented in fig. 1 show the absolute values of $\zeta$ potentials decrease with the decrease of PH. At $PH=2.7$, the $\zeta$ potentials go to zero, which is consistent to the experimental measurement of IEP~\cite{brian}. With the increase of the PH, the absolute values also increase. After experiencing a nearly plateau, the $\zeta$ potentials then start jumping to large absolute values at $PH=9.5$. Such $\zeta$ potential behaviors in the figure have captured the main feature of the relation between the $\zeta$ potentials and PH measured by experiments, except the latter shows an obvious linearity~\cite{brian}. The deviation of the calculated results away from the linearity is attributed to our neglect of the precise relation between the trap height and the local chemical environment at various PH and PC values. The study on the precise relation is not in the scope of this work, since we only concern about the mechanism of the charge neutralization at IEP.\\

\begin{figure}[!t]
\hspace*{-0.5cm}
\includegraphics [width=4.0in,height=3.0in]{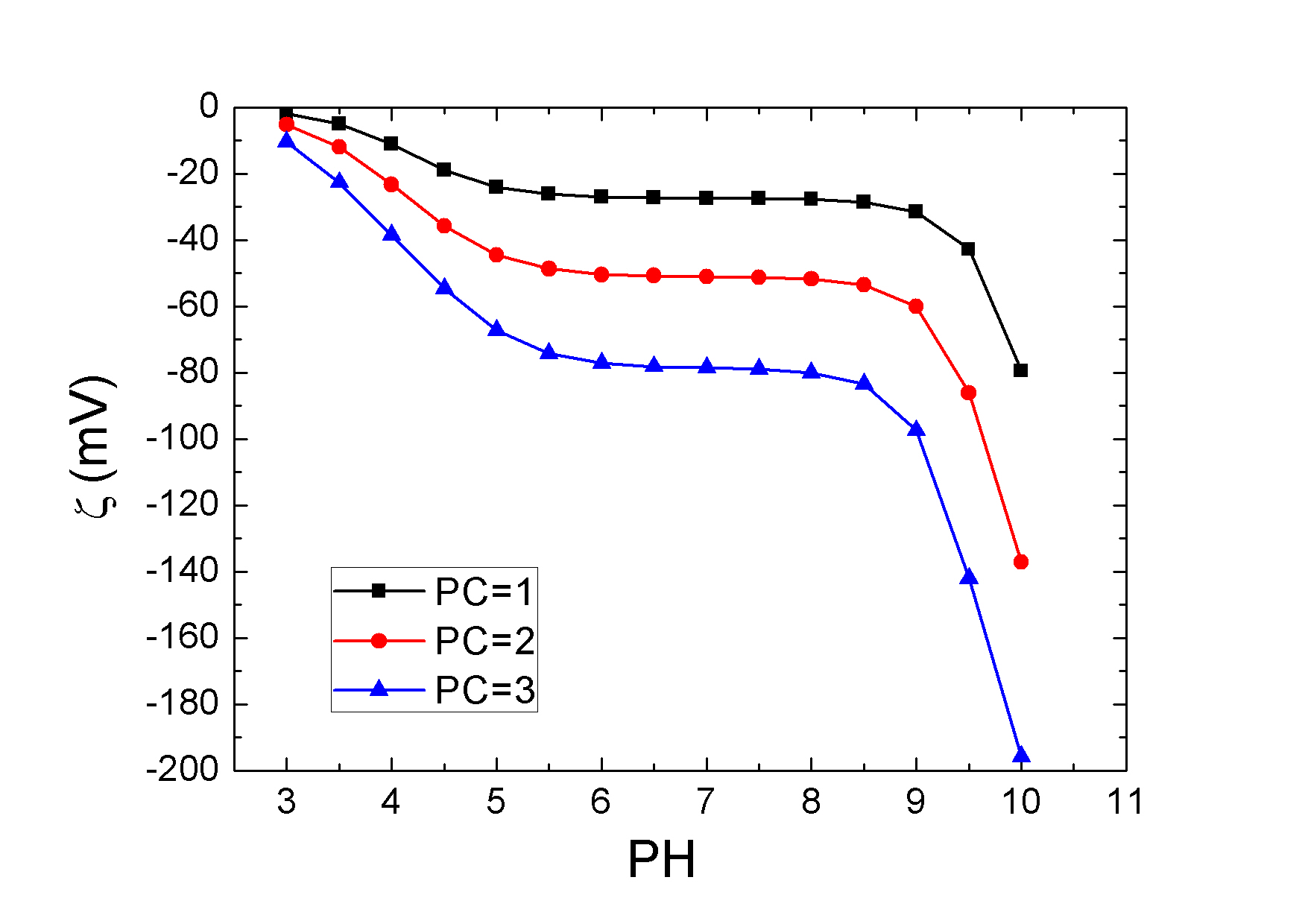}
\caption{$\zeta$ potential as a function of PH at PC=1,2,3. In the calculation, $R=20 \mu m$ and $\gamma=608 mV$ are used.}
\end{figure}

In order to reveal the origin of the results in fig. 1, the surface charge densities at $PC=1$ have been calculated in fig. 2(a). $\sigma$ representing the mount of $OH^-$ contributed from the water dissociation reaches its maximum at $PH=7$, and decreases when $PH$ is away from 7, no matter if the PH increases or decreases. The decrease of $\sigma$ at $PH \neq 7$ is due to the mass action law. At IEP, $\sigma$ is zero within the calculation error, meaning that enough high $H^+$ density at IEP suppresses the produce of $H^+$ from the water dissociation. $b$ has been referred to the $OH^-$ ion density in bulk limit trapped totally at the interface. logarithmic value of $b$ is observed to be linearly increasing with the increase of PH due to the increase of $OH^-$ ion density in bulk, and can be negligible when $PH <6$. The physical properties of the EDL basically is determined by the $s$. In the figure, it shows that the curve of $s$ overlaps that of $\sigma$ when $PH<7$ and meets the $b$ curve when $PH>9$. That means there exist two sources contributing to the charges in the stern layer. At $PH<6$, the charges in the stern layer are mainly originated from the dissociation of water molecules. And at $PH>9$, the charges in the stern layer are mainly from the bulk liquid. Since the $\sigma$ is zero at $PH=2.7$ due to the mass action law, no charges can be trapped in the stern layer, leading to the zero $\zeta$ potential at IEP. It also indicates in the fig. 1 that the $\zeta$ potential jumps to a large absolute value at $PH=9.5$ due to the existence of enough $OH^-$ ions in bulk liquid. What is more, it should be noted that the $s$ and $\zeta$ potential in the case of $PH>9.5$ can not go to infinity, and should be limited by two constraints. One constraint is due to the limited capacity of potential trap itself. The other is due to the limited lattices at the silica surface. The latter can also be understood as the steric effect. If the limitations are considered, the $s$ and $\zeta$ potentials will reach a plateau at $PH>9.5$. The detail calculation of the $s$ and $\zeta$ potential at a high PH is not the goal of this paper.\\

\begin{figure}[!t]
\hspace*{-0.5cm}
\includegraphics [width=4.0in,height=3.0in]{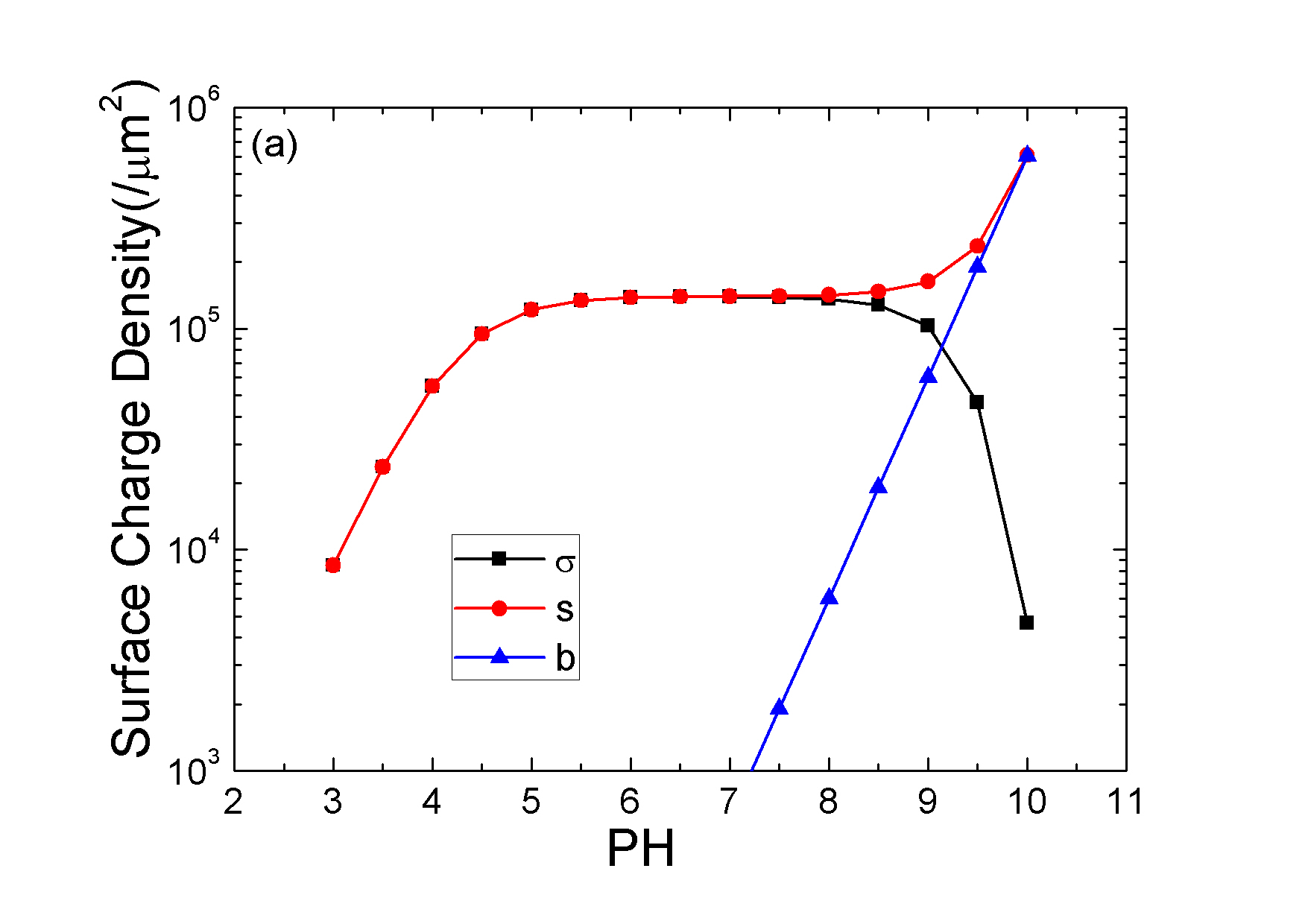}\\
\hspace*{-0.5cm}
\includegraphics [width=4.0in,height=3.0in]{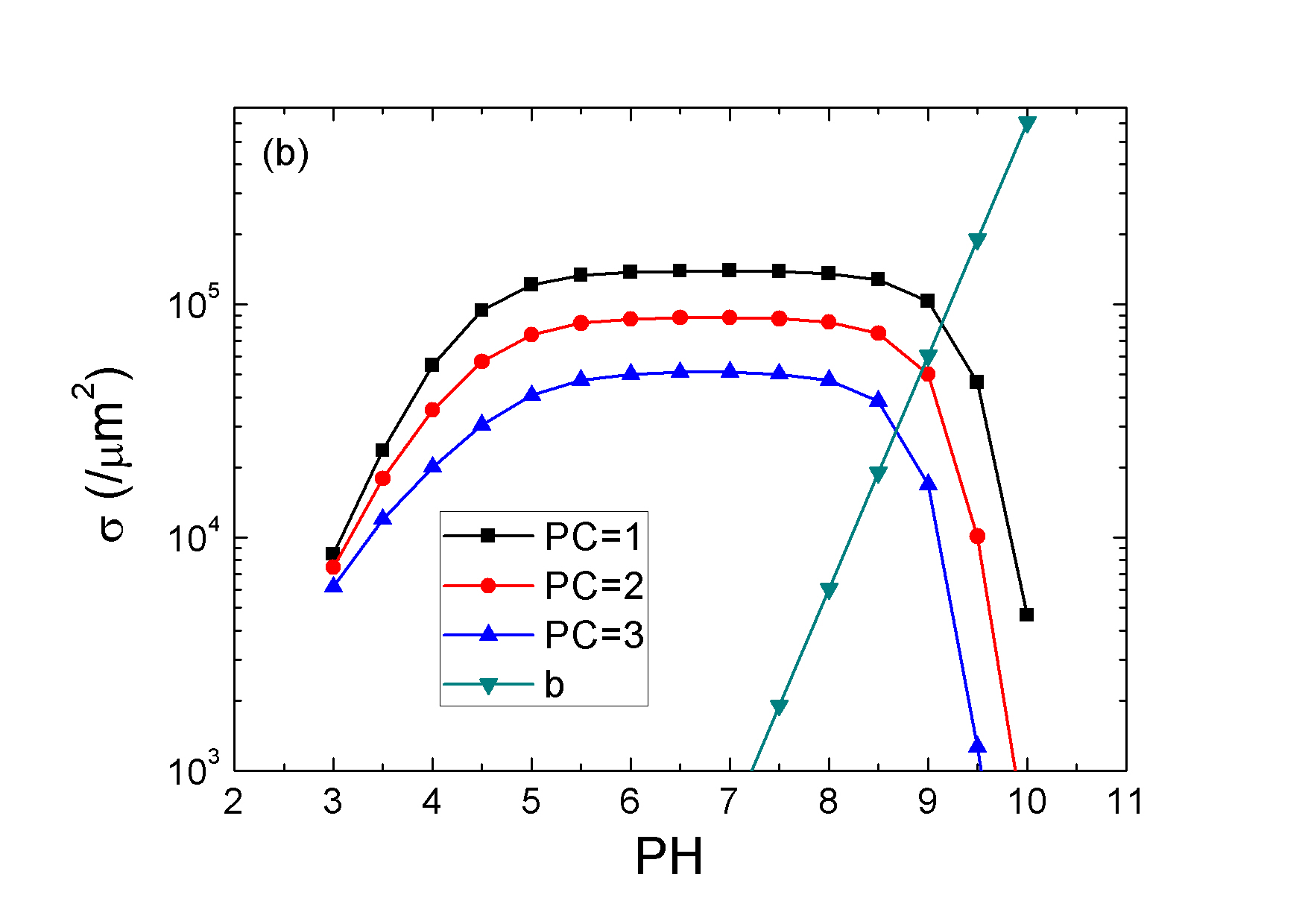}
\caption{Surface charge densities as functions of PH. In the calculation, $R=20 \mu m$ and $\gamma=608 mV$ are used. (a) surface charge densities of $\sigma$, $b$, and $s$ are presented for $PC=1$. (b) $\sigma$ is exhibited for PC=1,2,3 and the $b$ has also been presented for reference.}
\end{figure}

In the GC model, the occurrence of IEP is understood as the neutralization of the surface charges by $H^+$ ions from acid.  However, in the GC model, the source for the pre-existing surface charges has not been clarified. Even if the total $OH^-$ ions in the bulk is used as the source to be trapped, the surface charge density is not large enough to give an experimentally observable $\zeta$ potential, which has been confirmed by the $b$ curve in the fig. 2(a). Instead, the potential trap model can give the answer that the source for the surface charge at the PH close to IEP is from the dissociation of water molecules, and at the IEP the dissociation process is suppressed due to the mass action law.\\

\begin{figure}[!ht]
\hspace*{-0.5cm}
\includegraphics [width=4.0in,height=3.0in]{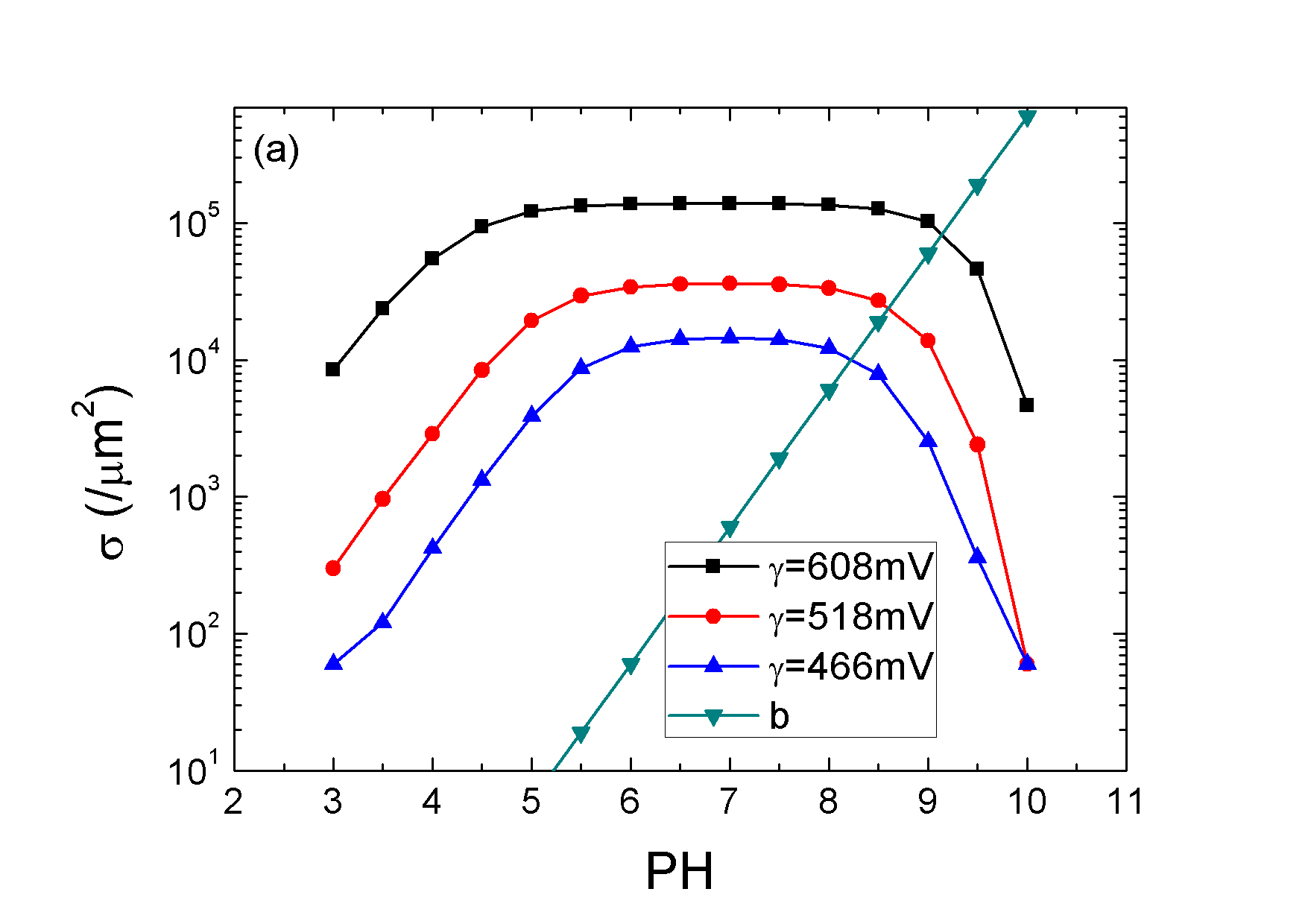}\\
\hspace*{-0.5cm}
\includegraphics [width=4.0in,height=3.0in]{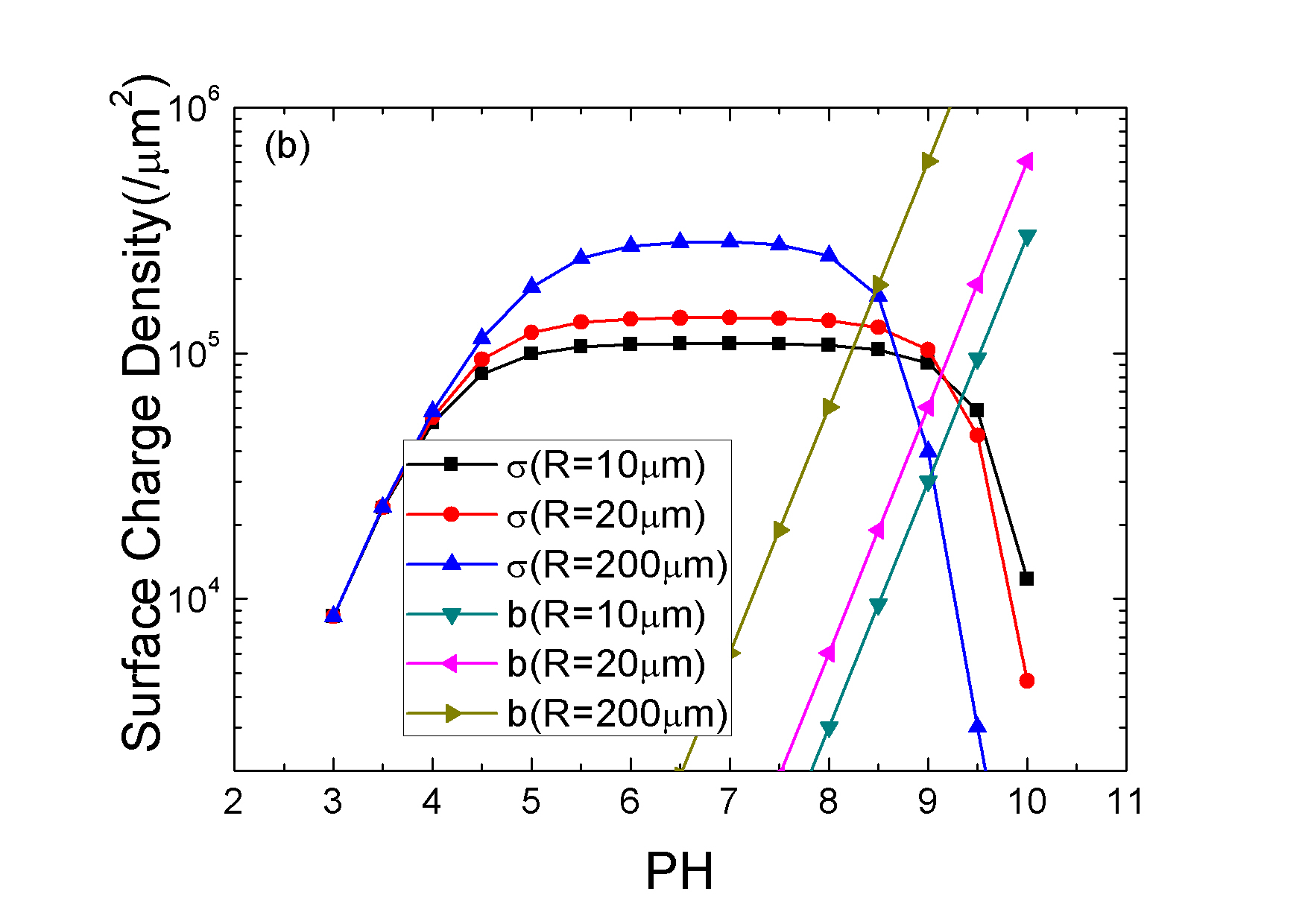}
\caption{Surface charge density as a function of PH at $PC=1$, (a) by varying $\gamma$ with $R=20\mu m$, (b) by varying R with $\gamma=608mV$.}
\end{figure}

Under the same condition of $\gamma=608mV$, the surface charge densities at various PC have been presented in fig.2(b). In the figure, the dissociation efficiency of water molecules reflected by $\sigma$ decreases with the increase of PC. That means salt concentration into the water can help to activate the water dissociation, but decrease the $\zeta$ potential (fig.1). Such phenomena can be understood as the following. Suppose a fixed charge density at the interface, the electric field magnitude at the slip plane can be obtained by using the Gaussian law. Such electric field would penetrate the Debye layer to the cylinder center, resulting in a potential drop. Since the potential has been fixed to be zero at the interface as the boundary condition, the potential at the cylinder center denoted as $\psi_c$ should be positive with the consideration that the Debye layer comprises of positive charges. The $\psi_c$ is the exact value equal to the absolute value of the $\zeta$ potential. It is well known that the add of salt into water can decrease the thickness of Debye layer. With the increase of the salt concentration in water, the decrease of the Debye layer will decrease the $\psi_c$ if the electric field is fixed at the slip plane as a constant before entering into the Debye layer. Then, according to the Boltzmann distribution, the decrease of positive $\psi_c$ will attract more positive charges in the Debye layer, which is realized by the increase of the surface negative charge density in the stern layer. Thus, the decrease of PC can increase $\sigma$, but the whole effect of the thickness decreasing of Debye layer results in the decrease of $\zeta$ potential.\\

It has also been found that $\sigma$ is strongly dependent on the trap height $\gamma$, shown in fig.3(a). A larger $\gamma$ can provide more energy for the dissociation of water molecules, leading to a larger $\sigma$. The relation between the $\sigma$ and the cylinder radius $R$ has also been presented in fig. 3(b). With the increase of $R$, more $OH^-$ ions in the bulk are provided as the source for the surface charge density, shifting the $b$ curve to a lower PH. And more neutral water molecules are provided for the dissociation with a larger $R$, which can shift the $\sigma$ curve upward to be larger. The relation between the $\zeta$ potential and $\gamma$ or $R$ can be found elsewhere~\cite{wan}.

\section{conclusion}
The potential trap model has been used to study the mechanism of the surface charge neutralization at IEP. It is found that at IEP the decrease of the charge density at the interface is due to the suppress of the water dissociation, instead of the neutralization by acid. The increase of the salt concentration in the water can increase the charge density in the stern layer, but decrease the $\zeta$ potential, which is consistent to the experimental results.

%\nocite{*}


\end{document}